\documentclass[letterpaper, 10 pt, journal, twoside]{IEEEtran}
\usepackage[pdftex]{graphicx} 
\usepackage{gensymb}
\usepackage{amsmath}
\usepackage{hyperref}
\IEEEoverridecommandlockouts  
\usepackage{booktabs}
\usepackage{xcolor}
\usepackage[normalem]{ulem}
\usepackage{tikz}
\usepackage{xfrac}
\usepackage{float}
\usepackage{placeins}
\usepackage[justification=centering]{caption}
 \usepackage[pscoord]{eso-pic}
  
\begin{document}

\title{Developing a Dual-Stage Vision Transformer Model for Lung Disease Classification}
\author{
\begin{tabular}{c} Anirudh Mazumder \\ Department of Engineering \\ University of North Texas \\ Denton, United States of America \\ anirudhmazumder@my.unt.edu\end{tabular} \and
\hspace{50pt} 
\begin{tabular}{c} Dr. Jianguo Liu \\ Department of Mathematics \\ University of North Texas \\ Denton, United States of America \\ jianguo.liu@unt.edu\end{tabular}
}
\maketitle
\begin{abstract}
Lung diseases have become a prevalent problem throughout the United States, affecting over 34 million people. Accurate and timely diagnosis of the different types of lung diseases is critical, and Artificial Intelligence (AI) methods could speed up these processes. A dual-stage vision transformer is built throughout this research by integrating a Vision Transformer (ViT) and a Swin Transformer to classify 14 different lung diseases from X-ray scans of patients with these diseases. The model showed promise when making predictions on an unseen testing subset of the dataset after data preprocessing and training the neural network. The model showed promise for accurately classifying lung diseases and diagnosing patients who suffer from these harmful diseases.
\end{abstract}
\section{Introduction}
Lung disease is a set of ailments, including COPD, asthma, and pneumonia, amongst many others, that prevent the lungs from performing their normal functions. Over 34 million people in the US have lung diseases \cite{yale2024pulmonologist}. In addition to the patients it affects, lung disease impacts the families of the patients as there are multiple mental and situational issues arising with receiving the burden of dealing with these lung diseases \cite{lungcarers}. The number of patients with lung disease is only increasing, in addition to the number of undiagnosed patients also increasing. Therefore, it is critical to develop solutions to help with the diagnosis and treatment of lung disease. Artificial Intelligence (AI) is trained on a large corpus of annotated images, aiming to learn about the underlying features of each image \cite{fi15110370}, \cite{Cahyo2023EarlyDetection}. AI models can predict if patients have lung diseases and their specific types of lung diseases. 
\section{Methodology}
\subsection{Dataset}
\begin{figure}[!htb]
	\centering
	\includegraphics[width=\columnwidth]{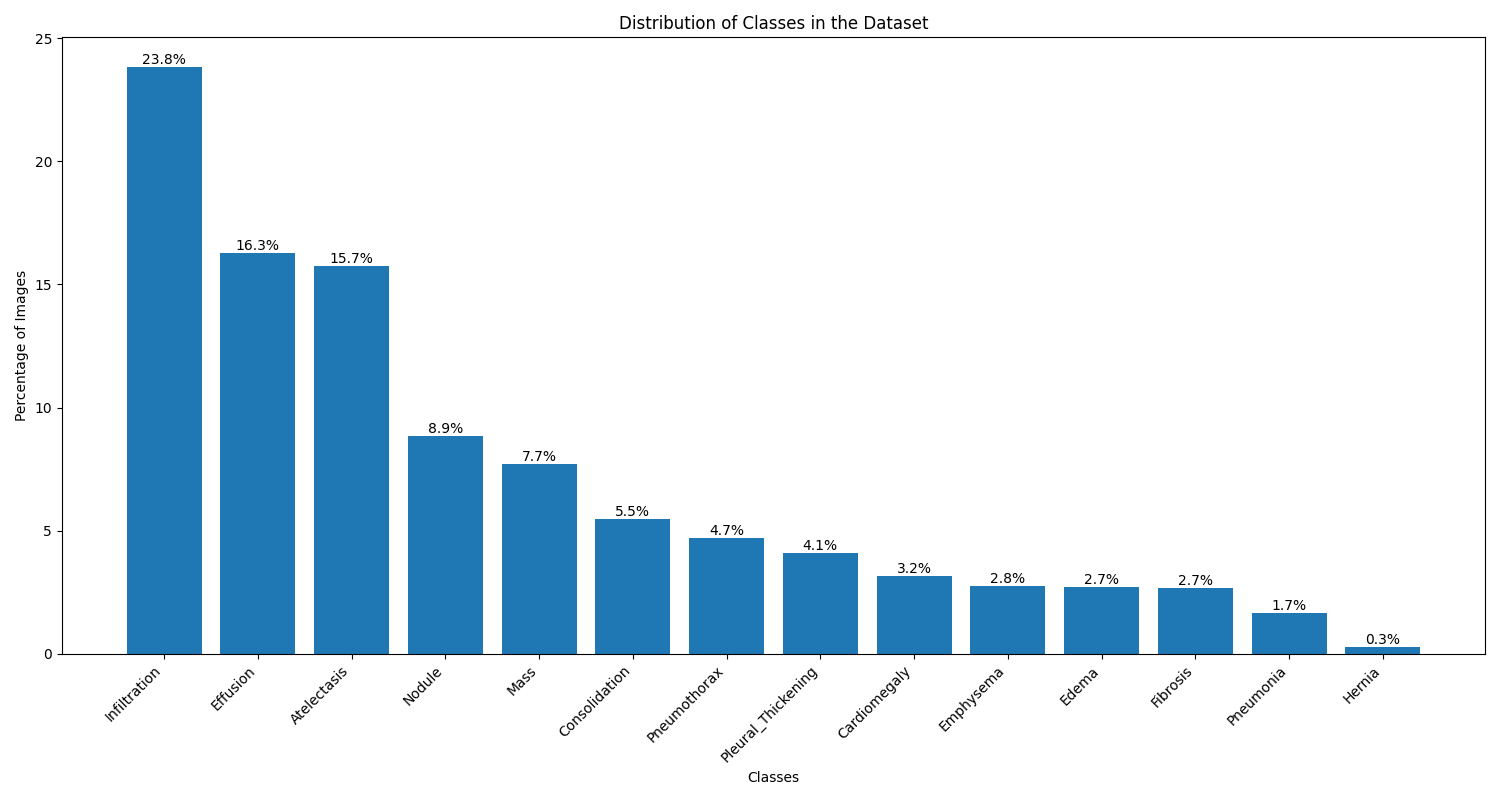}
	\caption{Distribution of the classes throughout the dataset.}
	\label{fig: Figure 2}
\end{figure}

The dataset that was utilized to develop the model can be found at \cite{wang2017chestxray}. The dataset contains fourteen different types of lung diseases or classes, including atelectasis, consolidation, infiltration, pneumothorax, edema, and more. With many classes, a critical piece of information that needed to be extracted from the dataset was the distribution of classes. Understanding how the classes are distributed is important because it can impact the model's performance and ability to understand certain lung diseases. The distribution can be seen in Figure \ref{fig: Figure 2}. Based on Figure \ref{fig: Figure 2}, it can be seen that there is a vast difference between some of the classes, indicating that there is a chance that the neural network underperforms on those under-stacked classes because they are not as common as some of the other features. 
\FloatBarrier
\FloatBarrier
\subsection{Data Preprocessing}
\subsubsection{Normalization}
The first preprocessing step was to normalize the data throughout the dataset. First, the images were resized to 224 x 224 pixels to normalize the data. Then, the images were normalized across the RGB color channels in both the mean and standard deviations of the pixels inside the images. The image was normalized into a range of [-1, 1], making it easier for a neural network to learn patterns on the image.
\subsubsection{Data Augmentation}
One other key step that was undertaken during image preprocessing was data augmentation. Data augmentation was performed by randomly flipping some of the images horizontally, increasing the diversity of the dataset. These transformations are important for the development of the model because they mitigate the model's likelihood of overfitting by developing variability in the training data.
\subsection{Model Architecture}
The model that was developed integrates a ViT and Swin Transformer, in a dual-stage approach to classify these lung disease images.
\subsubsection{Vision Transformer Component}
The ViT extracts the first set of features that the model learns from the images. The ViT model pre-trained on ImageNet-21k with a resolution of 224 x 224 pixels was used to define the architecture of this component. This architecture captures intricate patterns throughout an image by handing image patches as input tokens. In order to extract the features from the images, the model divides the images into non-overlapping patches and projects each patch into an embedding space while being processed through a series of transformer layers. The transformer layer gives high-dimensional representations of the images, which are pooled to produce a feature vector. Additionally, the ViT was developed such that it only had feature extraction capabilities and lacked classification capabilities. Once the feature vector is extracted, a global average pooling is applied to the feature vector because it averages the feature map across its spatial dimensions, giving a single feature vector summarizing an image \cite{10190115}. 
\subsubsection{Swin Transformer Component}
For the next component, a Swin Transformer was used to refine the features that were learned from the ViT by integrating a pre-trained Swin Transformer trained on ImageNet-1k at resolution 224 x 224 pixels. The Swin Transformer utilizes a hierarchical design and window-based attention mechanism to obtain general and contextual information about specific tasks \cite{9710580}. Similar to the ViT, the Swin Transformer was implemented such that it could impact the feature maps using its hierarchical attention layers, and the features were aggregated and reduced using the global average pooling function. 
\subsubsection{Feature Fusion}
In order to integrate the learned features from the ViT and Swin Transformer, their features were concatenated. To perform this concatenation, the ViT feature maps were resized to match the dimensions of the Swin Transformer's feature maps. After combining these features, both of the transformer models are processed using a fully connected layer, which gives the final classification results for the 14 lung disease classes. This is done using a linear connection layer, which has the dimensions of the combined feature size.
\subsubsection{Training and Optimization}
The loss function used while training the neural network was a Binary Cross-Entropy with Logits Loss Function, which has been proven to be effective at teaching neural networks to learn multiclass classification problems by teaching them to assign higher probabilities to the correct class labels for specific inputs \cite{pmlr-v202-mao23b}, \cite{ruby2020binary}. The Adam optimizer was used while training the model because it handles gradients and noisy data. The training loop followed the following training loop. First, the forward pass mechanism uses the ViT to predict the images. Next, the loss is calculated based on the difference between the predicted and actual labels, which integrates the Binary Cross-Entropy Loss with a Sigmoid layer into a single class. Then, a backward pass is used where the gradients of the loss are calculated with respect to the model parameters through backpropagation. Following the backpropagation, the parameters of the model are updated.
\section{Results}
\subsection{Model Performance}
In order to measure the performance of the model, a few different metrics were calculated. The first metric used to assess the model's performance on the dataset is a label-level accuracy metric. The model achieved an accuracy of 92.06\% on the label-level; however, the accuracy is skewed due to a class imbalance potentially making the predictions easier for the model. The accuracy metric shows that the model learned some features which are relevant for making predictions on patient's X-ray scans; however, it might perform poorly on underrepresented classes, thus with more fine-tuning and training the neural network would be able to make better predictions.

\begin{figure}[!htb]
	\centering
	\includegraphics[width=\columnwidth]{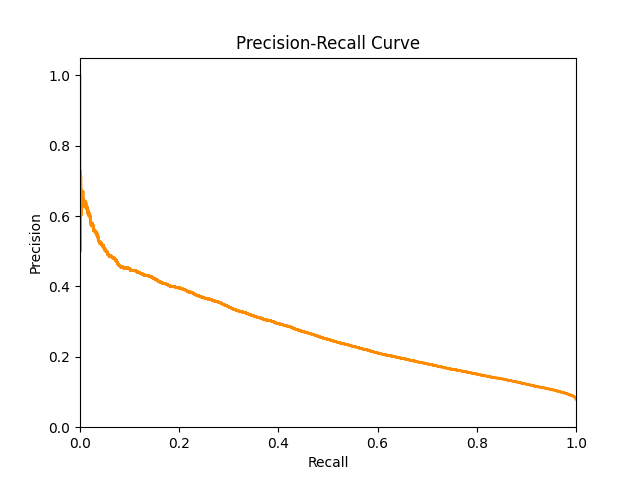}
	\caption{Graph showing the PR curve that was developed based on the model to assess its performance across different classification thresholds.}
	\label{fig: Figure 6}
\end{figure}

The last metric that was calculated for the model is a Precision-Recall (PR) Curve, which is used for understand the trade-off between precision and recall values across different classification thresholds the neural network could use. The precision metric shows the ratio of true positives to the sum of true and false positives. Additionally, the recall formula is defined as the ratio of true positives to the sum of true positives and false negatives. The PR curve can be seen in Figure \ref{fig: Figure 6}. The extracted PR curve demonstrates a downward trend, which indicates a trade-off between precision and recall as the threshold varies. Specifically, this trend suggests that as the model attempts to identify more positive instances, the precision decreases, indicating that the model's ability to correctly identify positive instances is compromised when increasing the number of detected positives. This downward trajectory maybe due to challenges in maintaining a balance between precision and recall, pointing to potential areas where the model could benefit from further tuning or improvements.
\FloatBarrier
\FloatBarrier
\section{Discussion}
\subsection{Conclusion}
This research implements a new neural network architecture to predict lung disease types based on X-ray scans of patients with different potential lung diseases. The neural network differentiates itself from prior methods by integrating two pre-trained vision transformer models to create a more holistic classification, using a dual-stage approach rather than integrating single vision transformers, CNNs, or hybrid models \cite{app13095521}. The neural network achieved an accuracy of 92.06\%, showing that it was able to learn the essential features required to classify these X-ray scans; however would require more fine-tuning and training to perform better on this task, and potentially outperform current state-of-the-art models.
\subsection{Future Works}
\subsubsection{Increase Computational Power}
The first key future work for this study is to increase the computational power used to train the neural network, as while conducting the research we were limited to the Tesla 4 GPU. Due to the limitation the number of epochs used when training the neural network was limited, so increasing attributes such as training time and batch sizes have the potential to enable the neural network to perform better.
\subsubsection{Benchmark on More Datasets}
Another key future work would be to benchmark this neural network architecture on other large scale datasets such as the CheXpert Dataset developed by Stanford University researchers \cite{chexpert}. This would be able to show the potential of the model to be applied to classify lung diseases over current state of the art models. Furthermore, other datasets could be used as testing sets to assess the ability of the neural network to accurately classify lung diseases on completely unseen datasets. 
\subsubsection{Clinical Application}
Another key future work would be to look into how these predictive models could be applied into the clinical workflow impact patients lives, by speeding up decision times, ensuring patients get their treatment faster.
\section{Acknowledgment}
We would like to thank the University of North Texas for providing us with the resources and support to conduct this research.
\bibliographystyle{IEEEtran}


\end{document}